\documentclass{elsart}
\usepackage{epsfig}

\begin{document}

\begin{frontmatter}

\title{An improved unified solver for compressible and incompressible fluids 
involving free surfaces. II. Multi-time-step integration and applications}
\author{Masato Ida}
\address{Satellite Venture Business Laboratory, Gunma University, \\
1--5--1 Tenjin-cho, Kiryu-shi, Gunma 376--8515, Japan\\
E-mail: ida@vbl.gunma-u.ac.jp
}
\thanks{\textit{Present address}: Institute of Industrial Science, 
The University of Tokyo, 
4--6--1 Komaba, Meguro-Ku, Tokyo 153--8505, Japan. 
E-mail: ida@icebeer.iis.u-tokyo.ac.jp
}

\begin{abstract}
An improved numerical solver for the unified solution of compressible and 
incompressible fluids involving interfaces is proposed. The present method 
is based on the CIP-CUP (Cubic Interpolated Propagation / Combined, Unified 
Procedure) method, which is a pressure-based semi-implicit solver for the 
Euler equations of fluid flows. In Part I of this series of articles [M. 
Ida, Comput. Phys. Commun. 132 (2000) 44], we proposed an improved scheme 
for the convection terms in the equations, which allowed us discontinuous 
descriptions of the density interface by replacing the cubic interpolation 
function used in the CIP scheme with a quadratic extrapolation function only 
around the interface. In this paper, as Part II of this series, the 
multi-time-step integration technique is adapted to the CIP-CUP integration. 
Because the CIP-CUP treats different-nature components in the fluid 
equations separately, the adaptation of the technique is straightforward. 
This modification allows us flexible determinations of the time interval, 
which results in an efficient and accurate integration. Furthermore, some 
additional discussion on our methods is presented. Finally, the application 
results to composite flow problems such as compressible and incompressible 
Kelvin-Helmholtz instabilities and the dynamics of two acoustically coupled 
deformable bubbles in a viscous liquid are provided.
\end{abstract}

\begin{keyword}
Unified solution \sep CIP \sep Compressible fluid \sep Incompressible fluid 
\sep Multi-time-step integration \sep Extrapolation \sep Free-surface flow \sep 
Bubble dynamics
\PACS 02.70.Bf \sep 47.11.+j \sep 47.20.Ma \sep 43.25.Yw
\end{keyword}

\end{frontmatter}

\section{Introduction}
This series of articles presents an improved solver for a challenging 
problem, the unified solution of compressible and incompressible fluids. 
After Harlow and Amsden proposed the ICE algorithm as a fully implicit solver 
for fluid equations in a conservative form \cite{ref1}, some approaches for 
this purpose have been investigated, such as the pressure-based 
semi-implicit algorithms \cite{ref2,ref3,ref4}, approaches based on the 
asymptotic expansion with respect to the local Mach number \cite{ref5,ref6}, 
and the boundary condition capturing method that treats compressible and 
incompressible materials separately \cite{ref7}. (See also recent reviews 
\cite{ref8,ref9} for more details.) Among them, the CIP-CUP method, a 
pressure-based semi-implicit algorithm proposed by Yabe and Wang \cite{ref3}, 
has already been applied to many practical multi-material problems such as the 
laser machining of a metal plate \cite{ref10}, comet Shoemaker-Levy 9's 
collision with the planet Jupiter \cite{ref11}, the interaction of a shock wave 
and a liquid drop \cite{ref12}, and the milk-crown formation on a liquid 
surface \cite{ref13}, and has been proven to be an efficient and robust solver 
for the unified solution \cite{ref14,ref15,ref16,ref17}. In this method, the 
Euler (or the Navier-Stokes) equations for fluid flows in a non-conservative 
form are selected as the governing equations, and the convection terms in 
the equations are solved explicitly by the CIP method \cite{ref18}, while the 
acoustic terms are solved implicitly by the CUP method \cite{ref3}. In Part I 
of this series \cite{ref19}, an improved solver for the convection terms was 
constructed using both an interpolation and an extrapolation function to 
describe the spatial profile of the density of materials. In the improved 
method, the cubic interpolation function used in the CIP scheme is replaced 
with the quadratic extrapolation functions constructed under some 
constraints for guaranteeing stability, only in a cell containing an 
interface between materials. This method allows us to solve the density 
interface with no dissipation across the interface, and is applicable to 
compressible flow problems, i.e., to variable-density problems.

As Part II of this series, this paper presents some further improvements for 
the CIP-CUP method, and also gives some application results. 
In Sec. \ref{sec3}, we 
attempt to incorporate the concept of the multi-time-step (MTS) integration 
techniques into the CIP-CUP algorithm. The MTS integration techniques, which 
allow us the efficient integration of dynamical systems involving some 
different time scales, have been used to solve a variety of problems, such 
as gravitational $N$-body problems \cite{ref20,ref21}, molecular dynamics 
\cite{ref22,ref23,ref24}, and atmospheric flow problems \cite{ref25}. This 
technique reduces the computational 
effort by integrating only rapidly varying components of a system (e.g., 
strongly accelerated particles or waves having fast phase speeds) with a 
small time interval, and others with a larger time interval. In the present 
study, based on this concept, different-nature terms in the fluid equations 
(convection, acoustic, and other terms) are solved using different time 
intervals. This modification allows us a more flexible determination of the 
time interval, and improves the accuracy and efficiency of the CIP-CUP. As 
will be shown, the adaptation of the MTS concept to the CIP-CUP integration 
is straightforward, because the CIP-CUP separately treats the terms having 
different natures, i.e., those having different characteristic time scales.

Section \ref{sec4} presents a modified scheme to update the spatial derivatives 
during the computational steps except for the convection parts. As was shown 
in Part I, the spatial derivatives of the dependent variables are used 
explicitly as additional dependent variables to solve the convection terms. 
In the convection process, the derivatives are updated to obey the equations 
derived by taking derivatives of the governing equations \cite{ref18,ref19}. 
Those spatial derivatives should be updated by some way also in the 
non-convection processes. In Refs.~\cite{ref18,ref26}, simple methods for this 
purpose have been 
proposed using classical centered differences. As was discussed in Part I, 
however, the use of such a classical discretization gives rise to the 
numerical dispersion and dissipation around the interfaces at which the 
spatial derivatives of the dependent variables tend to be discontinuous. We 
modify the centered scheme by adopting a simple extrapolation.

In Sec. \ref{sec5}, the applicability of the present methods to the 
compressible 
Navier-Stokes equations with a surface tension term is investigated, and in 
the Appendixes, some additional discussions regarding the averaging of the 
density (needed to solve the terms other than the convection) and the 
treatment of the surface-tension term are given. Section \ref{sec6} presents 
some application results for the compressible and incompressible 
Kelvin-Helmholtz instabilities and the multibubble dynamics in an acoustic 
field, and Sec. \ref{sec7} presents concluding remarks.

\section{The CUP method and its variants}

\subsection{The CUP method}
The CIP-CUP method \cite{ref3} is a semi-implicit solver for the Euler 
equations of fluid flows:
\begin{equation}
\label{eq1}
{\frac{{\partial \rho} }{{\partial t}}} + {\bf u} \cdot \nabla \rho = 
- \rho \nabla \cdot {\bf u} ,
\end{equation}

\begin{equation}
\label{eq2}
{\frac{{\partial {\bf u}}}{{\partial t}}} + {\bf u} \cdot \nabla 
{\bf u} = - {\frac{{\nabla p}}{{\rho} }} + {\frac{{{\bf F}}}{{\rho} }} ,
\end{equation}

\begin{equation}
\label{eq3}
{\frac{{\partial p}}{{\partial t}}} + {\bf u} \cdot \nabla p = - \rho 
C_{S}^{2} \nabla \cdot {\bf u} ,
\end{equation}
where $\rho $, ${\bf u}$, $p$, and $C_{S} $ denote the density, the 
velocity vector, the pressure, and the local sound speed, respectively, and 
{\bf F} may contain the 
viscosity, the surface tension, and external forces. This method separately 
solves the terms in Eqs. (\ref{eq1})--(\ref{eq3}) of different natures by a 
time splitting technique. The convection parts of these equations,
\[
\frac{{\partial \rho }}{{\partial t}} + {\bf u} \cdot \nabla \rho = 0 ,
\]

\[
\frac{{\partial {\bf u}}}{{\partial t}} + {\bf u} \cdot \nabla {\bf u} = 0 ,
\]

\[
\frac{{\partial p}}{{\partial t}} + {\bf u} \cdot \nabla p = 0 ,
\]
are solved by the CIP method \cite{ref18} (In the present work, of course, the 
hybrid interpolation-extrapolation method proposed in Part I is adapted), 
while the acoustic parts,
\begin{equation}
\label{eq4}
\frac{{\partial \rho }}{{\partial t}} =  - \rho \nabla  \cdot {\bf u} ,
\end{equation}

\begin{equation}
\label{eq5}
\frac{{\partial {\bf u}}}{{\partial t}} =  - \frac{{\nabla p}}{\rho } ,
\end{equation}

\begin{equation}
\label{eq6}
\frac{{\partial p}}{{\partial t}} =  - \rho C_S^2 \nabla  \cdot {\bf u} ,
\end{equation}
are solved by the CUP method, which is an implicit finite difference method. 
The remaining part,
\begin{equation}
\label{eq7}
\frac{{\partial {\bf u}}}{{\partial t}} = \frac{{\bf F}}{\rho } ,
\end{equation}
may be solved by some existing methods, such as the finite difference or 
finite volume method. (We refer to Eq. (\ref{eq7}) as the ``additional part''.) 
Solving these parts successively completes one step of the CIP-CUP time 
integration. In the following, the CUP method for the acoustic parts is 
concretely reviewed.

Discretizing the time derivatives on the LHS of Eqs. (\ref{eq4})--(\ref{eq6}) 
and estimating the spatial derivatives on the RHS with future values, one 
obtains
\begin{equation}
\label{eq8}
\frac{{\rho ^{n + 1}  - \rho ^* }}{{\Delta t}} =  - \rho ^* \nabla  \cdot 
{\bf u}^{n + 1} ,
\end{equation}
\begin{equation}
\label{eq9}
\frac{{{\bf u}^{n + 1}  - {\bf u}^* }}{{\Delta t}} =  
- \frac{{\nabla p^{n + 1} }}{{\rho ^* }},
\end{equation}
\begin{equation}
\label{eq10}
\frac{{p^{n + 1}  - p^* }}{{\Delta t}} =  - \rho ^* C_S^{*\,2} \nabla  \cdot 
{\bf u}^{n + 1} ,
\end{equation}
where $n$ is the number of the time step, $\Delta t$ is the time interval, 
and the quantities with the superscript $\ast$ indicate the 
values after solving the parts other than the acoustic parts. (The acoustic 
parts should be solved at the final stage of a time step, because, as is 
shown below, solving these parts enforces the divergence-free condition for 
an incompressible fluid.) Taking divergence of Eq. (\ref{eq9}) and substituting 
it into Eq. (\ref{eq10}) yield the following pressure equation:
\begin{equation}
\label{eq11}
\frac{{p^{n + 1}  - p^* }}{{\Delta t}} = \rho ^* C_S^{*\,2} \Delta t\,\nabla 
\cdot \frac{{\nabla p^{n + 1} }}{{\rho ^* }} - \rho ^* C_S^{*\,2} \nabla \cdot 
{\bf u}^* .
\end{equation}
(Substituting Eq. (\ref{eq6}) into this and rewriting as
\[
{{\left( {\frac{{p^{n + 1}  - p^* }}{{\Delta t}} - \frac{{\partial p^* }}
{{\partial t}}} \right)} \mathord{\left/ 
{\vphantom {{\left( {\frac{{p^{n + 1}  - p^* }}{{\Delta t}} 
- \frac{{\partial p^* }}{{\partial t}}} \right)} {\Delta t}}} \right.
 \kern-\nulldelimiterspace} {\Delta t}} = \rho ^* C_S^{*\,2} \nabla \cdot 
\frac{{\nabla p^{n + 1} }}{{\rho ^* }}
\]
finds that Eq. (\ref{eq11}) corresponds to a first-order approximation of the 
wave equation in terms of the pressure.) Almost the same pressure equation is 
given in Ref. \cite{ref4}, but discretization is done by a finite element 
technique.

After solving Eq. (\ref{eq11}) to get $p^{n + 1}$, the velocity is updated 
explicitly by Eq. (\ref{eq9}). Also, the density 
is updated explicitly by
\begin{equation}
\label{eq12}
\rho ^{n + 1}  = \rho ^*  + \frac{{p^{n + 1}  - p^* }}{{C_S^{*\,2} }} ,
\end{equation}
given by Eqs. (\ref{eq8}) and (\ref{eq10}). Generally, the spatial 
discretization for the 
above equations is performed using the 2nd-order centered finite 
differencing on the staggered grids.

When the sound speed is infinite, the pressure equation (\ref{eq11}) is reduced 
to an elliptic equation,

\begin{equation}
\label{eq13}
\nabla  \cdot \frac{{\nabla p^{n + 1} }}{{\rho ^* }} = \frac{{\nabla 
\cdot {\bf u}^* }}{{\Delta t}} ,
\end{equation}
which corresponds to that used in the SMAC algorithm \cite{ref27} for 
incompressible 
flows. This result reveals that the CUP method is applicable to both 
compressible and incompressible fluids.

In the case where the CIP or its variant is used to solve the convection 
parts, we also need to update the spatial derivatives, used explicitly in 
those schemes, in the non-convection processes. Simple schemes for this 
purpose have already been proposed \cite{ref18,ref26}. For a two-dimensional 
case, the scheme is represented as
\begin{equation}
\label{eq14}
\partial _x f_{i,j}^{n + 1}  - \partial _x f_{i,j}^*  = 
\frac{{f_{i + 1,j}^{n + 1}  - f_{i - 1,j}^{n + 1} }}{{2h}} 
- \frac{{f_{i + 1,j}^*  - f_{i - 1,j}^* }}{{2h}} ,
\end{equation}
\begin{equation}
\label{eq15}
\partial _y f_{i,j}^{n + 1}  - \partial _y f_{i,j}^*  = 
\frac{{f_{i,j + 1}^{n + 1}  - f_{i,j - 1}^{n + 1} }}{{2h}} 
- \frac{{f_{i,j + 1}^*  - f_{i,j - 1}^* }}{{2h}} ,
\end{equation}
where $f$ indicates an dependent variable, 
$\partial _x f = \partial f/\partial x$, 
$\partial _y f = \partial f/\partial y$, the subscripts $i$ and $j$ indicate 
$(x,y) = (i\,h,j\,h)$, and $h$ is the grid spacing 
(assumed to be constant for simplicity). In the case where the cross 
derivative $\partial _{xy} f$ is used in the convection process 
\cite{ref26,ref19}, the following equation is additionally used:
\begin{eqnarray}
\label{eq16}
\partial _{xy} f_{i,j}^{n + 1}  - \partial _{xy} f_{i,j}^*  = \frac{{f_{i + 1,j + 1}^{n + 1}  - f_{i - 1,j + 1}^{n + 1}  - f_{i + 1,j - 1}^{n + 1}  + f_{i - 1,j - 1}^{n + 1} }}{{4h^2 }} \nonumber \\
- \frac{{f_{i + 1,j + 1}^*  - f_{i - 1,j + 1}^*  - f_{i + 1,j - 1}^*  + f_{i - 1,j - 1}^* }}{{4h^2 }}.
\end{eqnarray}
Equations (\ref{eq14})--(\ref{eq16}) can be solved explicitly using the 
quantities obtained before and after solving the non-convection parts.

\subsection{Improved variants of the pressure equation}
\label{sec2:2}
In 1994 \cite{ref28}, Ito proposed an improved variant of the pressure equation 
[Eq. (\ref{eq11})] by incorporating the concept of the exponential method for a 
heat-conduction equation \cite{ref29}. The 1-D formula of the variant is 
represented by
\begin{equation}
\label{eq17}
\frac{{p^{n + 1}  - p^* }}{{\Delta t}} = \rho ^* C_S^{*\,2} \Delta t\,[\alpha \left( {\frac{{p_x^{n + 1} }}{{\rho ^* }}} \right)_x  + (1 - \alpha )\left( {\frac{{p_x^* }}{{\rho ^* }}} \right)_x ] - \rho ^* C_S^{*\,2} u_x^*  ,
\end{equation}
where $\alpha$ is the weighting factor determined theoretically as
\[
\alpha (E) = \frac{1}{{1 - \exp ( - E)}} - \frac{1}{E} ,
\]
\[
E = \left( {\frac{1}{{\rho _{i + 1/2}^* }} + \frac{1}{{\rho _{i - 1/2}^* }}} \right)\rho _i^* \left( {\frac{{{C_S}^{*} _{i} \Delta t}}{h}} \right)^2 ,
\]
and $\rho _{i \pm 1/2}^*$ is the density at $x = h\,(i + 1/2)$, determined 
approximately with $\rho _i^*$ and $\rho _{i \pm 1}^*$ (see Appendix 
A). For $E \to \infty$ (i.e., ${C_S}^* _{i} \Delta t/h \to \infty$), 
$\alpha$ converges to 1 and Eq. (\ref{eq17}) is reduced to the elliptic 
equation (\ref{eq13}); therefore, this variant is applicable to incompressible 
flows. For $E \to 0$, $\alpha$ becomes $1/2$, resulting in a higher resolution 
than that of the conventional CUP \cite{ref28,ref30}. We adopted this variant 
in the present study.

The 2-D formula of this variant (not shown in Ito's paper) may be represented by

\[
\frac{{p^{n + 1} - p^* }}{{\Delta t}} = \rho ^* C_S^{*\,2} \Delta t\,(\tilde G1 + \tilde G2) - \rho ^* C_S^{*\,2} \nabla \cdot {\bf u}^* ,
\]

where

\[
\tilde G1 = \alpha (E1)\left( {\frac{{p_x^{n + 1} }}{{\rho ^* }}} \right)_x  + (1 - \alpha (E1))\left( {\frac{{p_x^* }}{{\rho ^* }}} \right)_x ,
\]

\[
\tilde G2 = \alpha (E2)\left( {\frac{{p_y^{n + 1} }}{{\rho ^* }}} \right)_y  + (1 - \alpha (E2))\left( {\frac{{p_y^* }}{{\rho ^* }}} \right)_y ,
\]

\[
E1 = \left( {\frac{1}{{\rho _{i + 1/2,j}^* }} + \frac{1}{{\rho _{i - 1/2,j}^* }}} \right)\rho _{i,j}^* \left( {\frac{{{C_S}^* _{i,j} \Delta t}}{h}} \right)^2 ,
\]

\[
E2 = \left( {\frac{1}{{\rho _{i,j + 1/2}^* }} 
+ \frac{1}{{\rho _{i,j - 1/2}^* }}} \right)\rho _{i,j}^* \left( {\frac{{{C_S}^* _{i,j} \Delta t}}{h}} \right)^2 .
\]
We simplify this as follows to reduce the computational efforts:

\begin{eqnarray}
\label{eq18}
\frac{{p^{n + 1}  - p^* }}{{\Delta t}} =&& \rho ^* C_S^{*\,2} \Delta t\,[\alpha (E_{\max } )\nabla  \cdot \frac{{\nabla p^{n + 1} }}{{\rho ^* }} + (1 - \alpha (E_{\max } ))\nabla  \cdot \frac{{\nabla p^* }}{{\rho ^* }}] \nonumber \\
&&- \rho ^* C_S^{*\,2} \nabla  \cdot {\bf u}^* ,
\end{eqnarray}
where

\[
E_{\max }  = \max (E1,E2) .
\]
This formula requires only one weighting factor. This simplification may be 
valid because $E1 \approx E2$ generally.

In Ref. \cite{ref30}, an alternative variant was proposed. Though this variant 
can provide higher resolution of the sound wave than that obtained by Ito's 
variant, it is likely that the use of this variant as a part of the solver 
for fluid flows sometimes causes great violation of mass conservation 
\cite{ref31}. We therefore did not adapt it.

\section{Multi-time-step integration}
\label{sec3}
In many fluid flow problems, the time scale of a significant phenomenon is 
determined by the flow velocity (i.e., the characteristic speed of the 
convection parts), rather than by the sound speed or other characteristic 
speed. In certain cases, however, the acoustic parts (or others) have much 
smaller time scales (i.e., much greater characteristic speeds), resulting in 
an excessive restriction on the time interval, $\Delta t$. The CIP-CUP 
overcomes this problem by solving 
the acoustic parts implicitly. Such an approach, however, guarantees only 
the stability, and probably provides inaccurate results in certain 
situations. For example, under an intermediate condition of flows such as 
weakly compressible flows, the acoustic parts may need to be solved by using 
a time interval of sufficiently small $\kappa _C $ that makes $\kappa _u $ 
very small, since the compressibility described by 
the acoustic parts plays an important role, where

\[
\kappa _C  \equiv \max (C_S )\Delta t/h \quad 
{\rm and} \quad \kappa _u  \equiv \max (\left| {\bf u} \right|)\Delta t/h .
\]

To achieve an efficient computation even in such a situation, we adopt the 
multi-time-step (MTS) integration technique to the CIP-CUP method.

The adaptation of the MTS integration to the CIP-CUP is straightforward 
because the components of different time scales (convection, sound 
propagation, and others) are treated separately. The time propagation 
performed in the CIP-CUP can be represented schematically as follows:

\begin{equation}
\label{eq19}
{\bf U}^{**}  = L1(\Delta t)\,{\bf U}^n ,
\end{equation}

\begin{equation}
\label{eq20}
{\bf U}^*  = L2(\Delta t)\,{\bf U}^{**} ,
\end{equation}

\begin{equation}
\label{eq21}
{\bf U}^{n + 1}  = L3(\Delta t)\,{\bf U}^* ,
\end{equation}
where ${\bf U} = (\rho ,{\bf u},p)$ and the operators $L1(\Delta t)$, 
$L2(\Delta t)$, and $L3(\Delta t)$ indicate the 
discrete propagators for the corresponding integration steps (the 
convection, the additional, and the acoustic steps, respectively). These 
equations can be summarized as

\begin{equation}
\label{eq22}
{\bf U}^{n + 1}  = L3(\Delta t)L2(\Delta t)L1(\Delta t)\,{\bf U}^n .
\end{equation}
If we factor $L2(\Delta t)$ and $L3(\Delta t)$ into some identical 
pieces, we get

\begin{equation}
\label{eq23}
{\bf U}^{n + 1}  = [L3(\Delta t/m3)]^{m3} [L2(\Delta t/m2)]^{m2} L1(\Delta t)\,{\bf U}^n ,
\end{equation}
where $m2$ and $m3$ are positive 
integers. This equation means that, after solving the convection parts with 
$\Delta t$, the additional part is solved $m2$ times with $\Delta t/m2$, and 
subsequently the acoustic parts are solved $m3$ times with $\Delta t/m3$. 
If necessary, we divide $L2$ into two parts as $L2 = L2^{(2)} L2^{(1)}$, and 
modify Eq. (\ref{eq23}) as
\begin{equation}
\label{eq24}
{\bf U}^{n + 1}  = [L3(\Delta t/m3)\,L2^{(2)} (\Delta t/m3)]^{m3} \,[L2^{(1)} (\Delta t/m2)]^{m2} L1(\Delta t)\,{\bf U}^n 
\end{equation}
or
\begin{eqnarray}
\label{eq25}
{\bf U}^{n + 1}  = [L3(\Delta t/m3)\,L2^{(2)} (\Delta t/m3)\,[L2^{(1)} (\Delta t/(m2\,m3))]^{m2} ]^{m3} L1(\Delta t)\,{\bf U}^n . \nonumber \\
{}
\end{eqnarray}
Such a treatment of the additional parts is necessary, e.g., when the 
surface-tension term exists. (The surface-tension part, represented by 
$L2^{(2)}$, should be solved 
together with the acoustic parts because the surface tension always needs to 
balance with the pressure jump at the interface.) Some other forms of 
factorization adjusted for a governing system can be employed.

We determine the fundamental time interval, $\Delta t$, by

\[
\Delta t = \min \left( {c1\frac{h}{{\max (\left| {\bf u} \right|)}},c2\frac{h}{{\max (C_S )}}} \right) ,
\]
where the parameters are typically set to $c1 = 0.2$ and $c2 = 10$. $m3$ is 
determined so that, e.g., the following condition is satisfied:

\[
\frac{{\Delta t}}{{m3}} \le \min \left( {c3\frac{h}{{\max (\left| {{C_S} _{i + 1,j}  - {C_S} _{i,j} } \right|)}},c3\frac{h}{{\max (\left| {{C_S} _{i,j + 1}  - {C_S} _{i,j} } \right|)}},\Delta t} \right) ,
\]
where the spatial difference of sound speed is used as a criterion. If the 
surface-tension term exists, the stability condition needed to solve it 
\cite{ref32,ref33},

\begin{equation}
\label{eq26}
\frac{{\Delta t}}{{m3}} \le \Delta t_{st}  \equiv \sqrt {\frac{{(\rho _a  + \rho _b )h^3 }}{{4\pi \sigma }}} ,
\end{equation}
should be taken into consideration to determine $m3$, where $\rho _a$ and 
$\rho _b$ are the densities of materials on different sides of 
the interface, and $\sigma$ 
is the surface-tension coefficient. Also, $m2$ is determined based on the 
stability condition of a method for the viscous term or others.

\section{Stabilization of the derivative advancement}
\label{sec4}
At the interfaces where the value or derivatives of physical quantities are 
in general discontinuous, the use of the conventional centered schemes 
(\ref{eq14})--(\ref{eq16}) for updating the spatial derivatives is not suitable, providing less accurate results, as will be demonstrated below using 
a numerical example. To modify the schemes, we recall here the extrapolation 
concept investigated in Part I \cite{ref19} and by others 
\cite{ref34,ref35,ref36,ref37}. As has been pointed out in Ref. \cite{ref19}, 
an interpolation or differencing across the phase boundary 
may not physically valid, and gives rise to serious numerical errors. In the 
following, those schemes are modified by adapting a simple extrapolation.

Equation (\ref{eq14}) can be rewritten as

\begin{equation}
\label{eq27}
\partial _x f_{i,j}^{n + 1}  - \partial _x f_{i,j}^*  = \frac{1}{2}\left( {\frac{{d_{i + 1,j} - d_{i,j} }}{h} + \frac{{d_{i,j} - d_{i - 1,j} }}{h}} \right) ,
\end{equation}

\[
d_{i,j} \equiv f_{i,j}^{n + 1}  - f_{i,j}^* .
\]
We introduce here a switching parameter $H$, defined as

\begin{equation}
\label{eq28}
H_{i + 1/2,j}  = \left\{ {\begin{array}{*{20}c}
   {1  \quad} \hfill & {{\rm for}\quad \phi _{i + 1,j}^{n + 1}  \cdot \phi _{i,j}^{n + 1}  > 0,} \hfill  \\
   0 \hfill & {{\rm otherwise,}} \hfill  \\
\end{array}} \right.
\end{equation}

\begin{equation}
\label{eq29}
H_{i,j + 1/2}  = \left\{ {\begin{array}{*{20}c}
   {1 \quad} \hfill & {{\rm for}\quad \phi _{i,j + 1}^{n + 1}  \cdot \phi _{i,j}^{n + 1}  > 0,} \hfill  \\
   0 \hfill & {{\rm otherwise,}} \hfill  \\
\end{array}} \right.
\end{equation}
where $\phi$ is the ID function (a density function or a level set function) 
defined as $\phi > 0$ in the region 
occupied by a material and $\phi < 0$ elsewhere, and is updated by solving

\begin{equation}
\label{eq30}
\frac{{\partial \phi }}{{\partial t}} + {\bf u} \cdot \nabla \phi  = 0 .
\end{equation}
Using this, we modify Eq. (\ref{eq27}) as

\begin{equation}
\label{eq31}
\partial _x f_{i,j}^{n + 1}  - \partial _x f_{i,j}^*  = \frac{1}{2}\left( {H_{i + 1/2,j} \frac{{d_{i + 1,j} - d_{i,j} }}{h} + H_{i - 1/2,j} \frac{{d_{i,j}  - d_{i - 1,j} }}{h}} \right) .
\end{equation}
This modification vanishes the derivative $(d_{i + 1,j} - d_{i,j} )/h$ when 
$\phi _{i + 1,j}^{n + 1}  \cdot \phi _{i,j}^{n + 1}  < 0$, resulting in the 
advancement of $\partial _x f_{i,j}$ using only the values of an identical 
material. This procedure corresponds to an extrapolation such as 
$d_{i + 1,j} = d_{i,j}$. Equation (\ref{eq15}) can be modified in the 
same manner, as

\begin{equation}
\label{eq32}
\partial _y f_{i,j}^{n + 1}  - \partial _y f_{i,j}^*  = \frac{1}{2}\left( {H_{i,j + 1/2} \frac{{d_{i,j + 1}  - d_{i,j} }}{h} + H_{i,j - 1/2} \frac{{d_{i,j}  - d_{i,j - 1} }}{h}} \right) .
\end{equation}
The scheme (\ref{eq16}) for the cross derivative can be rewritten as

\begin{eqnarray}
\label{eq33}
\partial _{xy} f_{i,j}^{n + 1} - \partial _{xy} f_{i,j}^* = 
\frac{1}{{4h^2}}[(d_{i + 1,j + 1} - d_{i,j} ) - (d_{i - 1,j + 1} - d_{i,j} ) 
\nonumber \\
- (d_{i + 1,j - 1}  - d_{i,j} ) + (d_{i - 1,j - 1}  - d_{i,j} )].
\end{eqnarray}
We modify this as

\begin{eqnarray}
\label{eq34}
\partial _{xy} f_{i,j}^{n + 1} &&- \partial _{xy} f_{i,j}^* = \frac{1}{{4h^2 }} 
\nonumber \\ 
&&\times [H_{i + 1/2,j + 1/2} (d_{i + 1,j + 1} - d_{i,j}) - H_{i - 1/2,j + 1/2} (d_{i - 1,j + 1} - d_{i,j}) \nonumber \\
&&- H_{i + 1/2,j - 1/2} (d_{i + 1,j - 1} - d_{i,j}) + H_{i - 1/2,j - 1/2} (d_{i - 1,j - 1} - d_{i,j} )],
\end{eqnarray}
where

\begin{equation}
\label{eq35}
H_{i \pm 1/2,j \pm 1/2}  = \left\{ {\begin{array}{*{20}c}
   {1 \quad } \hfill & {{\rm for}\quad \phi _{i \pm 1,j \pm 1}^{n + 1}  \cdot \phi _{i,j}^{n + 1}  > 0,} \hfill  \\
   0 \hfill & {{\rm otherwise}.} \hfill  \\
\end{array}} \right.
\end{equation}

The extrapolations used above are merely rough compared with those used to 
solve the convection parts \cite{ref19,ref36}; these, however, might be 
sufficient, 
because the derivatives require lower accuracy than those required for the 
non-derivatives represented here by $f$.

\begin{figure}
\begin{center}
\leavevmode
\epsfxsize = 12 cm
\epsffile{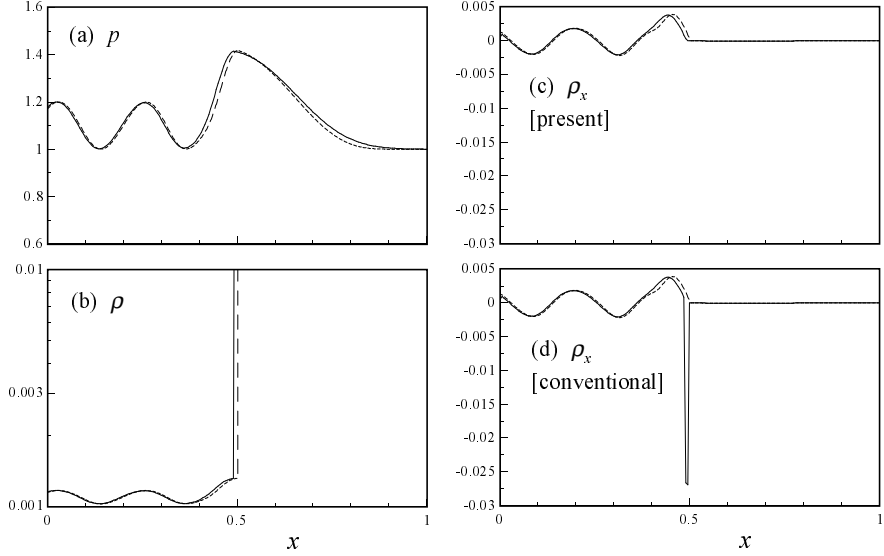}
\caption{
Nonlinear propagation of sound wave. Pressure (a), density (b), 
density gradient calculated by the present method (c), and by the 
conventional method (d), at $t = 1.602 \times 10^{ - 2} $. The dashed lines 
denote the result for $\Delta t/4$ and $h/4$.}
\label{fig1}
\end{center}
\end{figure}

Let us perform a numerical test to demonstrate the effectiveness of the 
above modification. A one-dimensional nonlinear sound propagation is solved 
based on Eqs. (\ref{eq4})--(\ref{eq6}). The convection and other components 
are neglected for simplicity. The initial condition is

\[
\begin{array}{l}
 \rho (x,0) = \left\{ {\begin{array}{*{20}c}
   {1.025 \times 10^{ - 3} } \hfill & {{\rm for}\quad x < 0.5,} \hfill  \\
   1 \hfill & {{\rm elsewhere},} \hfill  \\
\end{array}} \right. \\ 
 u(x,0) = 0,\quad p(x,0) = 1, \\ 
 x \in [0,\infty ), \\ 
 \end{array}
\]
and the boundary condition is

\[
p(0,t) = 1.1 - 0.1\cos \omega _1 t \quad {\rm and} \quad 
\partial u(0,t)/\partial x = 0,
\]
where the angular frequency $\omega _1$ is determined so that the wavelength of 
the emitted sound wave is $0.225$. The initial values 
of all derivatives are zero. The square of sound speed is determined by

\[
C_S^2  = \left\{ {\begin{array}{*{20}c}
   {7(p + 3172.04)/\rho } \hfill & {{\rm for}\quad x < 0.5,} \hfill  \\
   {1.4p/\rho } \hfill & {{\rm elsewhere}.} \hfill  \\
\end{array}} \right.
\]
Other parameters are $\Delta t = 2 \times 10^{ - 5}$ and 
$h = 5 \times 10^{ - 3}$. (Under this condition, $\kappa _C $ is about 0.6 for 
$x \ge 0.5$ and is about 0.15 elsewhere.) The ID function is set to be a color 
function,

\[
\phi  = \left\{ {\begin{array}{*{20}c}
   1 \hfill & {{\rm for}\quad x < 0.5,} \hfill  \\
   { - 1} \hfill & {{\rm elsewhere}.} \hfill  \\
\end{array}} \right.
\]
The matrix equation given by discretizing Eq. (\ref{eq18}) is solved by the 
Red-Black Gauss-Seidel method with the convergence criterion of 
$\varepsilon _{res}  < 10^{ - 6}$, where

\[
\varepsilon _{res}  = {{\left\| {p_i^{n + 1,(k)}  - p_i^{n + 1,(k - 1)} } \right\|_2 } \mathord{\left/
 {\vphantom {{\left\| {p_i^{n + 1,(k)}  - p_i^{n + 1,(k - 1)} } \right\|_2 } {\left\| {p_i^* } \right\|_2 }}} \right.
 \kern-\nulldelimiterspace} {\left\| {p_i^* } \right\|_2 }} ,
\]
and $k$ denotes the number of the Gauss-Seidel iteration. Figures \ref{fig1}(a) 
and \ref{fig1}(b) show the calculated pressure and density distributions at 
$t = 1.602 \times 10^{ - 2}$. Also, Figs. \ref{fig1}(c) and \ref{fig1}(d) show 
the density gradients at the same $t$, calculated using the conventional and 
the modified schemes. (In the acoustic step, the derivatives are not used to 
update the non-derivatives; thus, the results shown in Figs. \ref{fig1}(a) and 
\ref{fig1}(b) are irrespective of the schemes for the derivatives.) As can be 
clearly seen, the result obtained by the conventional scheme has a strong 
overshoot at the 
interface, while the result obtained by the modified scheme is smooth. The 
result given using the modified scheme with $\Delta t/4$ and $h/4$ (the dashed 
lines) is similar to the latter result. These results 
can roughly prove the validity of the modified scheme, although more 
detailed discussions should be necessary.

\section{Application to the compressible Navier-Stokes equations}
\label{sec5}
In this section, as a summary of methods, we illustrate how to apply the 
methods discussed in this series to the compressible Navier-Stokes equations 
with a surface tension term:

\begin{eqnarray}
\label{eq36}
\frac{{\partial \rho }}{{\partial \,t}} + {\bf u} \cdot \nabla \rho 
&=& - \rho \nabla  \cdot {\bf u}, \\
\label{eq37}
\frac{{\partial {\bf u}}}{{\partial t}} + {\bf u} \cdot \nabla {\bf u} 
&=& - \frac{{\nabla p}}{\rho } + \frac{1}{\rho }\left( {2\nabla  \cdot (\mu {\bf T}) - \frac{2}{3}\nabla (\mu \nabla  \cdot {\bf u})} \right) + \frac{{{\bf F}_{st} }}{\rho }, \\
\label{eq38}
\frac{{\partial p}}{{\partial \,t}} + {\bf u} \cdot \nabla p 
&=& - \rho C_s^2 \nabla  \cdot {\bf u},
\end{eqnarray}
where ${\bf T}$ is the deformation tensor, $\mu$ is the viscosity coefficient, 
and ${\bf F}_{st}$ indicates the surface tension as a volume force. We 
factor them into the following four systems:

\textit{Convection system:}

\[
\left\{ {\frac{{\partial \rho }}{{\partial \,t}} + {\bf u} \cdot \nabla \rho  = 0,\quad \frac{{\partial {\bf u}}}{{\partial \,t}} + {\bf u} \cdot \nabla {\bf u} = 0,\quad \frac{{\partial p}}{{\partial \,t}} + {\bf u} \cdot \nabla p = 0,} \right.
\]

\textit{Viscous system:}

\[
\left\{ {\frac{{\partial \rho }}{{\partial \,t}} = 0,\quad \frac{{\partial {\bf u}}}{{\partial \,t}} = \frac{1}{\rho }\left( {2\nabla  \cdot (\mu {\bf T}) - \frac{2}{3}\nabla (\mu \nabla  \cdot {\bf u})} \right),\quad \frac{{\partial p}}{{\partial \,t}} = 0,} \right.
\]

\textit{Surface-tension system:}

\[
\left\{ {\frac{{\partial \rho }}{{\partial \,t}} = 0,\quad \frac{{\partial {\bf u}}}{{\partial \,t}} = \frac{{{\bf F}_{st} }}{\rho },\quad \frac{{\partial p}}{{\partial \,t}} = 0,} \right.
\]

\textit{Acoustic system:}

\[
\left\{ {\frac{{\partial \rho }}{{\partial \,t}} =  - \rho \nabla  \cdot {\bf u},\quad \frac{{\partial {\bf u}}}{{\partial \,t}} =  - \frac{{\nabla p}}{\rho },\quad \frac{{\partial p}}{{\partial \,t}} =  - \rho C_s^2 \nabla  \cdot {\bf u}.} \right.
\]

Successively solving these four systems by the MTS technique completes one 
step of the time integration. Among the convection equations, one in terms 
of the density is solved by the hybrid interpolation-extrapolation method 
introduced in Part I, and the others are solved by the conventional CIP (the 
``Type B'' multidimensional formula \cite{ref26} is adapted). The time 
integration of $\phi$ is performed together with the convection system. The 
viscous system is solved by the conventional second-order centered finite 
differencing, and the surface-tension system is solved by the CSF model 
\cite{ref32}, coupled with the smoothing procedure introduced in Appendix B. 
The acoustic system is solved 
by the modified CUP method reviewed in Sec. \ref{sec2:2}.

\section{Application results}
\label{sec6}
This section presents the application results for the multiphase flow 
problems, given using the present methods. Compressible and incompressible 
Kelvin-Helmholtz instabilities and multibubble dynamics in an acoustic field 
were selected as application examples.

\begin{figure}
\begin{center}
\leavevmode
\epsfxsize = 10 cm
\epsffile{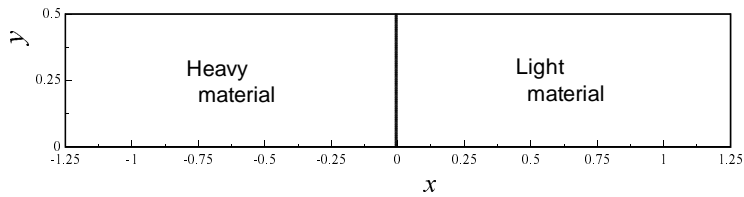}
\caption{
Initial material arrangement for Kelvin-Helmholtz instabilities.}
\label{fig2}
\end{center}
\end{figure}

\begin{figure}
\begin{center}
\leavevmode
\epsfxsize = 10 cm
\epsffile{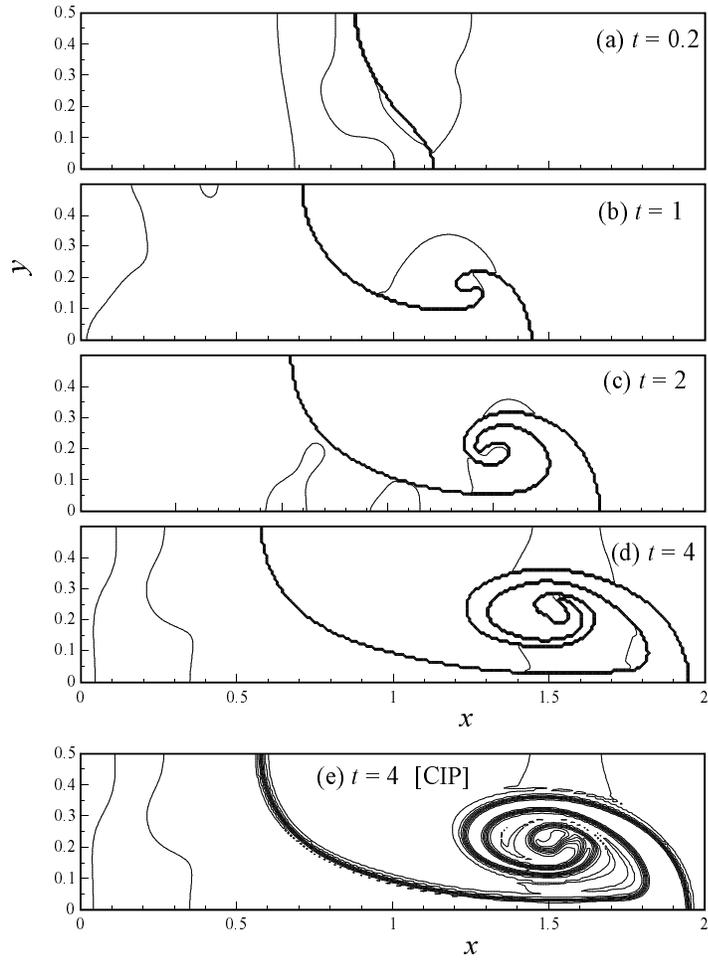}
\caption{
Results for compressible Kelvin-Helmholtz instability by the hybrid 
method (a--d) and by the conventional CIP (e).}
\label{fig3}
\end{center}
\end{figure}

\begin{figure}
\begin{center}
\leavevmode
\epsfxsize = 12 cm
\epsffile{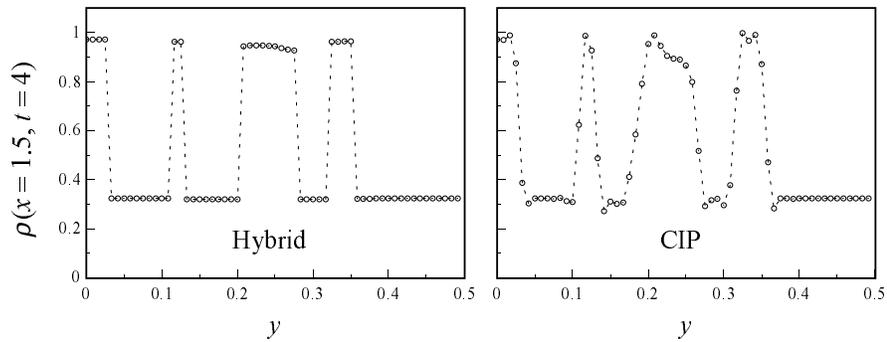}
\caption{
Density profiles on $x = 1.5$ at $t = 4$, by 
the hybrid method (left) and by the conventional CIP (right).}
\label{fig4}
\end{center}
\end{figure}

\begin{figure}
\begin{center}
\leavevmode
\epsfxsize = 10 cm
\epsffile{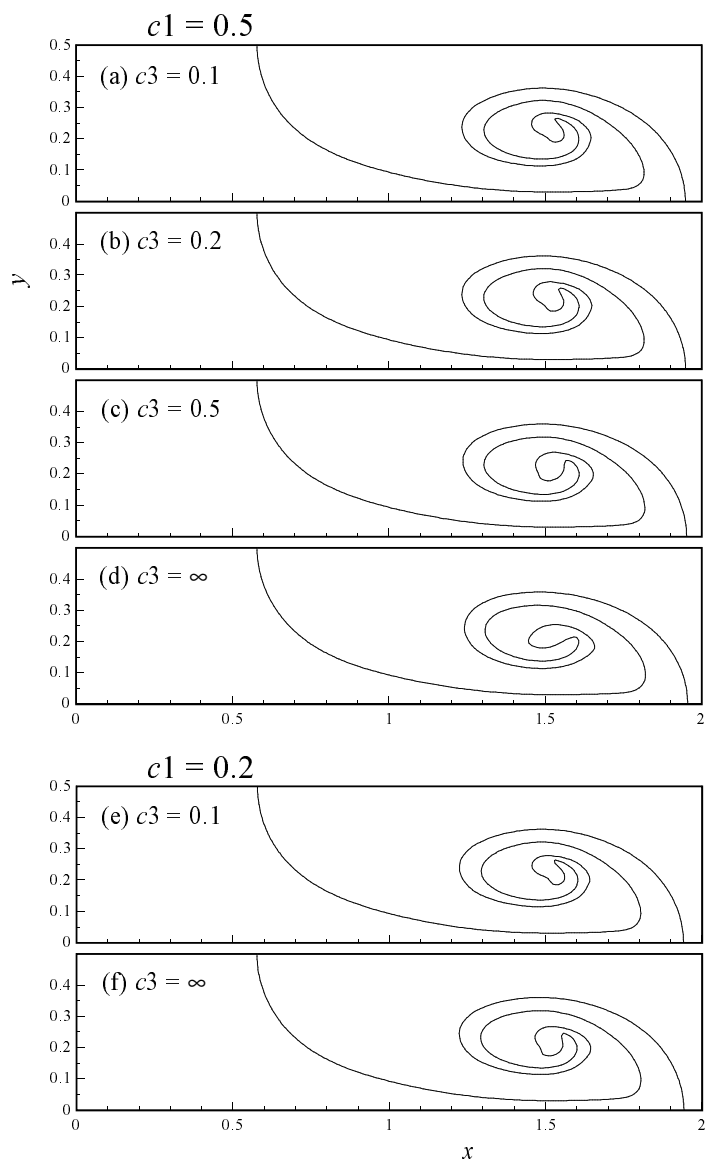}
\caption{
Interfaces ($\phi  = 0$ surfaces) at $t = 4$ for $c1 = 0.5$ (a--d) or 
$c1 = 0.2$ (e and f) and different $c3$.}
\label{fig5}
\end{center}
\end{figure}

\textbf{Compressible and incompressible Kelvin-Helmholtz instabilities.} It 
is well known that an interface between fluids of different densities is 
unstable. When the velocity field around the interface is perturbed, the 
interface forms a complicated structure \cite{ref38}. This instability, called 
the Kelvin-Helmholtz instability or the Rayleigh-Taylor instability, has been 
the subject of quite a number of theoretical, experimental, and numerical 
studies \cite{ref38,ref39,ref40,ref41,ref42,ref43,ref44,ref45}, and even the 
CIP have been employed to investigate it \cite{ref41,ref46}. Using this 
example, the effectiveness of the extrapolation technique for the convection 
parts and that of the MTS technique are demonstrated. The initial arrangement 
of materials is shown in Fig. \ref{fig2}. The 
density of the heavy material (one for $x < 0$) is $\rho _h  = 1$, and that of 
the light one $\rho _l  = 1/3$, and the initial pressure is 
$p_0 (x,y,t = 0) = 1$. The initial velocity field 
[${\bf u}(x,y,t = 0) = (u_0 (x,y),v_0 (x,y))$] is set to

\[
u_0 (x,y) = {\mathop{\rm sgn}} (x)\,u_{\max } \exp ( - 2\pi \left| x \right|/\lambda )\sin (2\pi y/\lambda ) ,
\]

\[
v_0 (x,y) = u_{\max } \exp ( - 2\pi \left| x \right|/\lambda )\cos (2\pi y/\lambda ) ,
\]

\[
(x,y) \in [ - 1.25,1.25] \times [0,0.5] ,
\]
which satisfies $\nabla  \cdot {\bf u} = 0$, where $u_{\max }$ is the 
maximum velocity and $\lambda$ ($ = 1$) is the 
wavelength of the perturbation. The non-dimensional parameter 
$\lambda \sqrt {\rho _h p_0 } /\mu$ is fixed to 1400, 
and $\mu$ is assumed to 
be constant. The surface tension and the gravity are neglected. The slip 
boundary condition is adapted to the boundaries of $y = 0$ and $y = 0.5$, and 
the Neumann boundary condition ($\partial /\partial \vec n = 0$) to 
$x =  - 1.25$ and $x = 1.25$. Figures \ref{fig3}(a)--\ref{fig3}(d) show the 
time sequence of 
$\rho$ for $u_{\max }  = 0.7$, $C_S^2  = 1.4p/\rho$ (the 
Mach number under this condition is about 0.6), and $h$ 
$( = \Delta x = \Delta y) = 0.5/60$, at $t = 4$. Here, the linear monotone 
function \cite{ref19} is used for the extrapolation, and Eq. (\ref{eq23}) is 
used for the MTS integration. The parameters 
for the time intervals are $c1 = 0.5$, $c2 = \infty$, and $c3 = 0.1$. (The 
inner iteration number for the acoustic parts, $m3$, is $7 \sim 13$ in this 
case.) These results present sharp descriptions of the 
density interface, while the numerical diffusion and oscillation can be 
observed in the result given when the convection parts are solved by the 
conventional CIP (Figs. \ref{fig3}(e) and \ref{fig4}). Figures 
\ref{fig5}(a)--\ref{fig5}(d) show 
$\phi  = 0$ surfaces at $t = 4$ for different $c3$. (For 
$c3 = \infty$, the MTS integration is disabled, that is, a single time step is 
used.) As is clearly seen, decreasing $c3$ changes the result. A similar result 
to that for $c3 = 0.1$ can be given, even using a smaller $\Delta t$ (see 
Figs. \ref{fig5}(e) and \ref{fig5}(f), which show the results for $c1 = 0.2$ 
and different $c3$). 
These results reveal that integrating the acoustic 
parts with a smaller time interval improves the accuracy of the solution, 
even if the time interval for the convection parts holds.

\begin{figure}
\begin{center}
\leavevmode
\epsfxsize = 10 cm
\epsffile{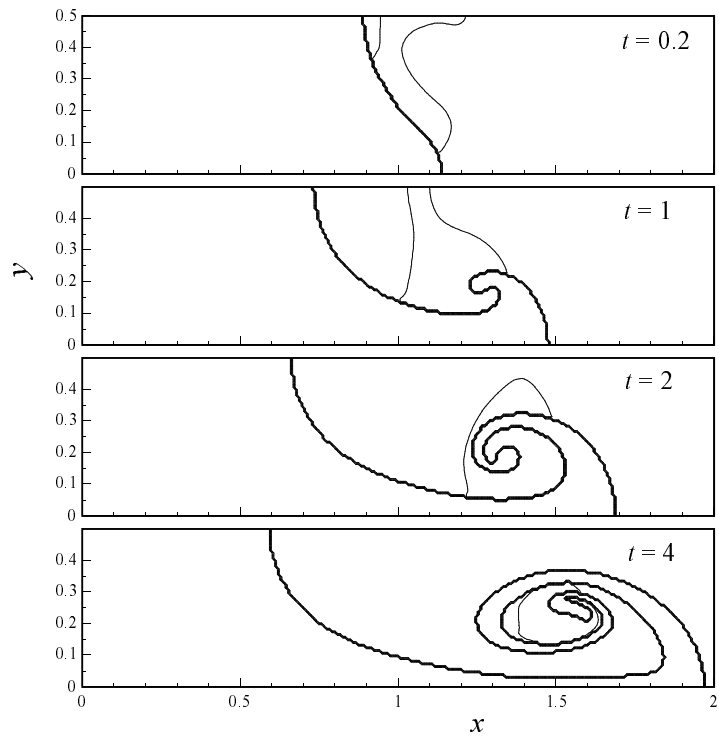}
\caption{
Kelvin-Helmholtz instability in a composite flow of compressible and 
incompressible fluids.}
\label{fig6}
\end{center}
\end{figure}

\begin{figure}
\begin{center}
\leavevmode
\epsfxsize = 10 cm
\epsffile{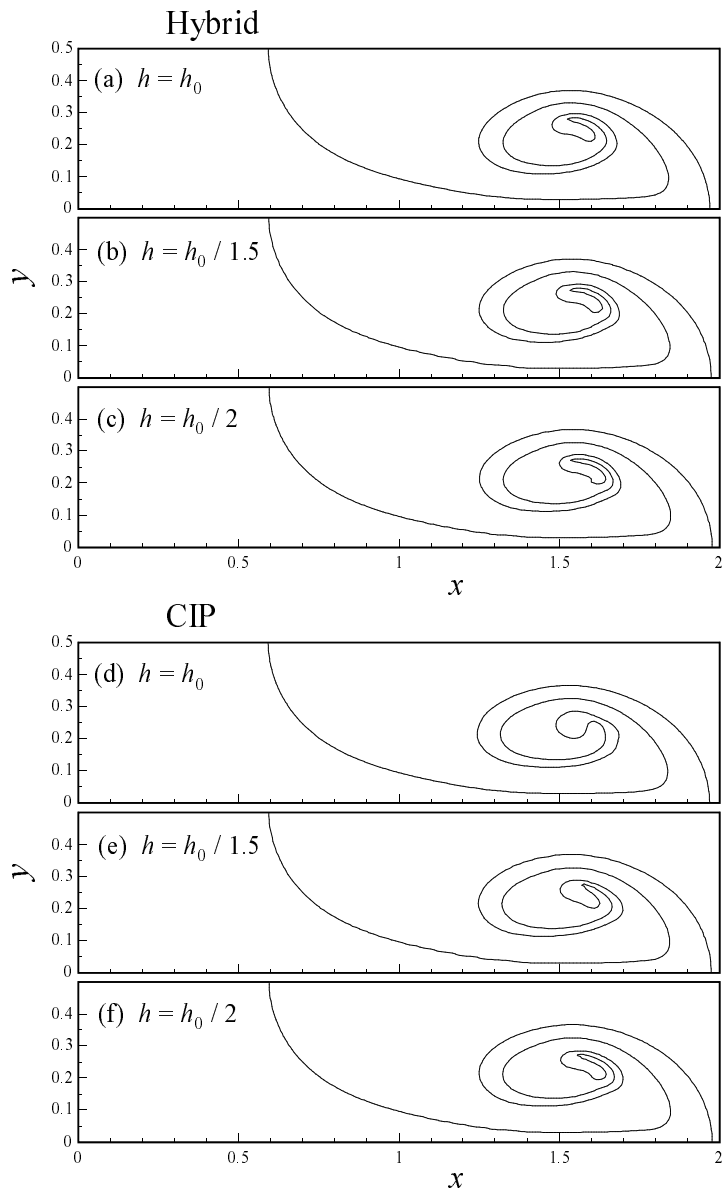}
\caption{
Convergency test. Results at $t = 4$ given by the hybrid method (a--c) 
and by the conventional CIP (d--f), for $h = h_0$, $h_0 /1.5$, and $h_0 /2$.}
\label{fig7}
\end{center}
\end{figure}

While the above example can be solved by an explicit scheme without any 
difficulty, the next example is problematical. Figure 
\ref{fig6} shows the results given by artificially resetting the sound speed to 
$C_S^2  = 1000 \times 1.4p/\rho$ for only the heavy 
material. This setting results in a much greater sound speed than the flow 
velocity, and also results in the coexistence of materials of greatly 
different compressibilities. The parameters for the time integration are 
$c1 = 0.25$, $c2 = \infty$, and $c3 = 5.0$. (Under this setting, 
$m3 = 3 \sim 5$ and $\kappa _C /m3 \approx 4 \sim 5$.) This result is not 
changed noticeably by decreasing the time interval, proving the applicability 
of the present method to a high $\kappa _C$ 
condition and proving its robustness. Using this example, we perform here a 
convergency test. Figures \ref{fig7}(a)--\ref{fig7}(c) show the interfaces at 
$t = 4$ given using 
the same parameters except for the grid width set to $h = h_0$, $h_0 /1.5$, or 
$h_0 /2$, respectively, where $h_0  = 0.5/60$. The refined results are in good 
agreement with that for $h = h_0$. In contrast, 
the result given by solving the convection parts with only the interpolation 
(i.e. by the conventional CIP) changes as the computational grids are 
refined (see Figs. \ref{fig7}(d)--\ref{fig7}(f)); the convergency of the 
solution is 
obviously lower than that obtained by the hybrid method. These results show 
that the improvement given in Part I accelerates the convergency, and that 
the hybrid scheme can accurately describe such deformable interfaces.

\textbf{Two-bubble dynamics in an acoustic field.} When a sound wave is 
applied, a bubble immersed in a liquid begins volume oscillation. When other 
bubbles exist, they interact acoustically with each other, resulting in the 
changes in the oscillation amplitude and phase and the effective resonance 
frequencies \cite{ref47,ref48,ref49}. Moreover, it is known that an acoustic 
interaction force called the secondary Bjerknes force acts between such 
pulsating bubbles \cite{ref50,ref51}. The force is attractive when the bubbles 
pulsate in-phase, 
and is repulsive otherwise. Such effects resulting from the radiative 
interaction between bubbles have been important subjects not only for 
physicists and mechanical and acoustical engineers, but also for medical 
engineers and chemists \cite{ref52,ref53,ref54,ref55,ref56}, and novel insights 
regarding these effects have been provided even in very recent years 
\cite{ref57,ref58,ref59,ref60,ref61}. (In Refs. \cite{ref59,ref60}, for 
example, the author discovered that in multibubble cases, the 
phase shift of the bubbles' pulsation can take place not only at their 
natural frequencies, but also at some other driving frequencies. In a more 
recent paper \cite{ref61}, it was found that the latter characteristic 
frequencies cause the sign reversal of the secondary Bjerknes force 
\cite{ref51}.)

Solving this problem numerically has some interesting points. (1) The 
gas inside the bubbles and the liquid surrounding them have quite different 
densities and compressibilities. (2) The compressibility of the gas is 
not negligible in principle, meaning that a solver for incompressible flows is 
not applicable. (3) Especially for small bubbles, the viscosity of the 
surrounding liquid and the surface tension are not negligible. (4) Many 
different time scales exist, determined by the flow velocity, the sound 
speed, the surface tension, the viscosity, the bubbles' natural frequencies, 
and the frequency of an external sound. (5) The topology of the interfaces 
changes when the bubbles coalesce as a result of their radiative 
interaction. Based on these interesting factors, we consider this problem to 
be a very good example for demonstrating the abilities of our methods. 

The axisymmetric coordinate $(r,z)$ is selected for the computational domain, 
and the centers of the 
bubbles are located on the central axis. Bubbles 1 and 2, whose equilibrium 
radii are $R_{10}  = 5$ $\mu$m and $R_{20}  = 9$ $\mu$m, respectively, are 
filled with a gas of the constant specific heat ratio $\gamma  = 1.33$. The 
equilibrium density and the viscosity 
coefficient are 1.226 kg/m$^3$ and $1.78 \times 10^{ - 5}$ Pa s 
for the gas phase, and 1000 kg/m$^3$ and $1.137 \times 10^{ - 3}$ Pa s for the 
liquid phase. The surface tension coefficient is 
$\sigma  = 7.28 \times 10^{ - 2}$ Pa m, and the equilibrium pressure of the 
liquid is $P_0  = 101.3$ kPa. The initial pressures inside the bubbles are 
thus determined by $P_0  + 2\sigma /R_{j0}$ 
(for $j = 1$ and $2$), in order to balance with the surface tension, where 
$2\sigma /R_{10} \approx 0.29P_0$ and $2\sigma /R_{20} \approx 0.16P_0$. 
The sound speed is set to $C_S^2  = \gamma p/\rho$ for the gas, and to 
$C_S^2  = 7(p + 3172.04P_0 )/\rho$ for the liquid. The external sound is 
applied as the boundary condition to the pressure by 
$p(\Gamma ) = P_0 (1 + 0.3\sin \omega t)$, where $\Gamma$ denotes the 
boundaries of the computational domain except for $r = 0$, and $\omega$ is the 
angular 
frequency of the sound. The boundary condition for the velocity is free. The 
initial distance between the centers of the bubbles [$D(t = 0)$] is fixed to 
$20$ $\mu$m. The grid width assumed to be constant is set to 
$h = (5/20)$ $\mu$m, and the total number of grids is $100 \times 310$. 
$\rho _a  + \rho _b$ in Eq. (\ref{eq26}) is set to 1000 
kg/m$^3$ (= the equilibrium density of the liquid), because the density of the 
gas is negligible, and that of the liquid can be assumed to be almost constant 
in time.

\begin{figure}
\begin{center}
\leavevmode
\epsfxsize = 12 cm
\epsffile{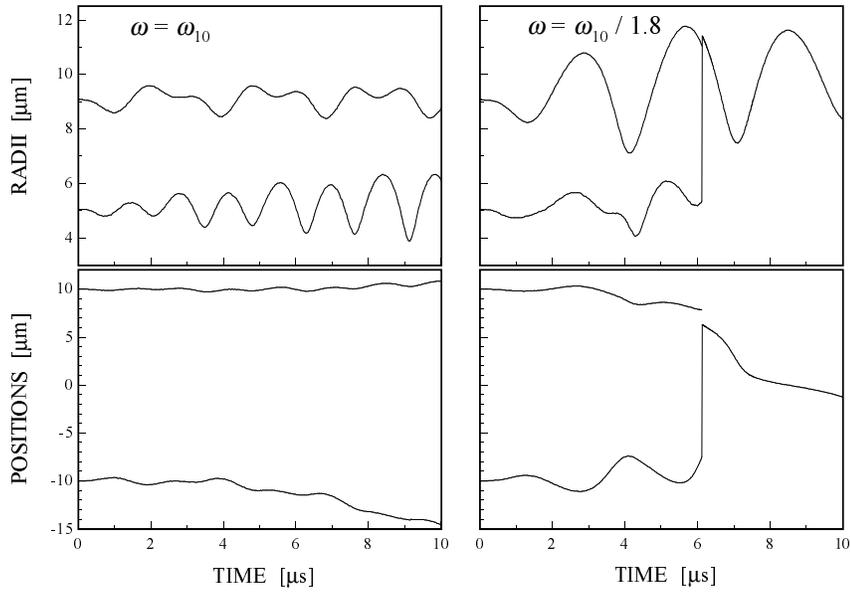}
\caption{
Bubbles' mean radii and positions (the lower one is for the smaller 
bubble) for $\omega  = \omega _{10}$ and $\omega  = \omega _{10} /1.8$ as 
functions of time. The coalescence of the bubbles occurred at the time when the 
number of the lines became one. }
\label{fig8}
\end{center}
\end{figure}

\begin{figure}
\begin{center}
\leavevmode
\epsfxsize = 10 cm
\epsffile{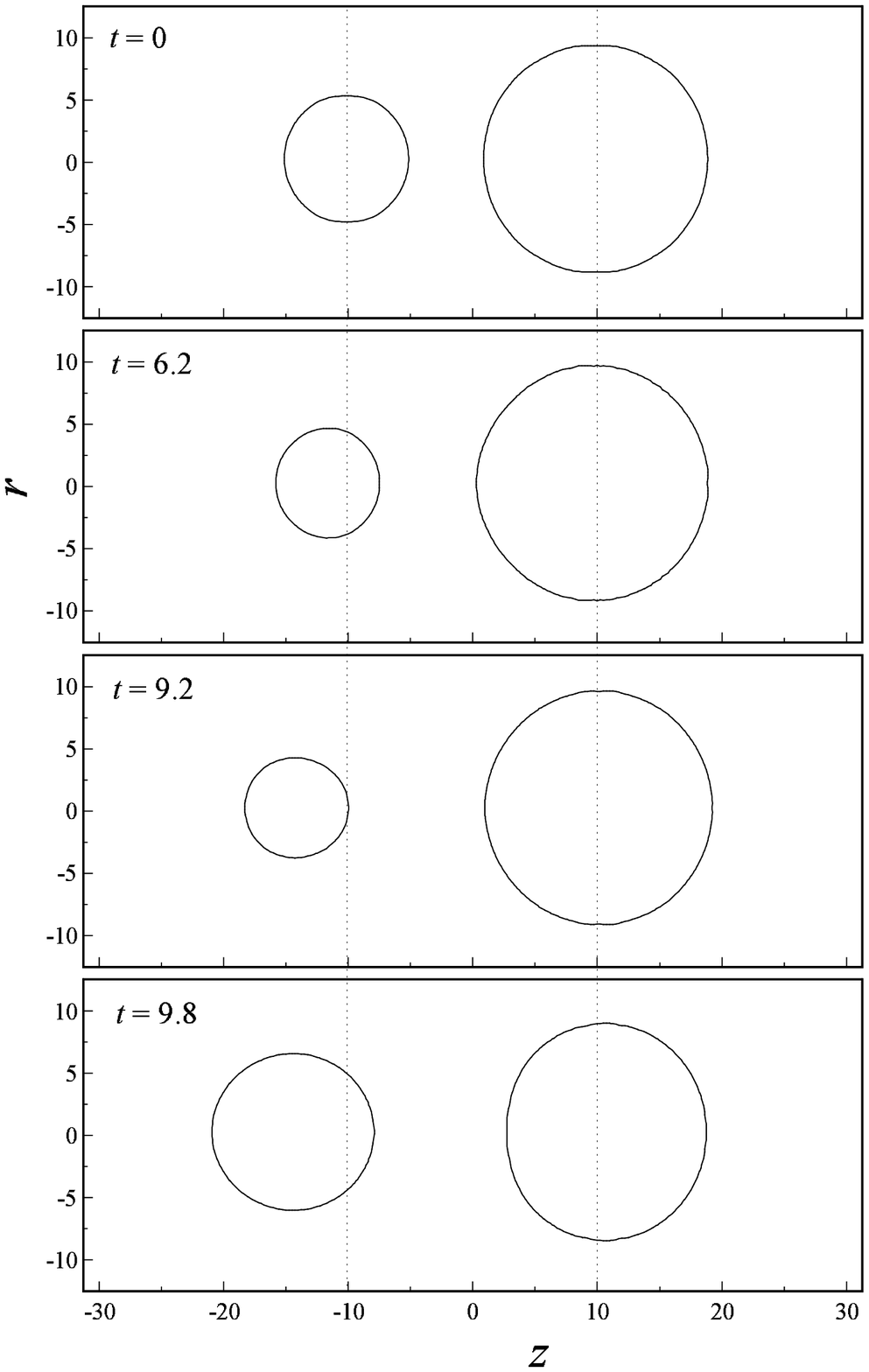}
\caption{
Calculated interfaces for $\omega  = \omega _{10}$ at selected times.}
\label{fig9}
\end{center}
\end{figure}

\begin{figure}
\begin{center}
\leavevmode
\epsfxsize = 10 cm
\epsffile{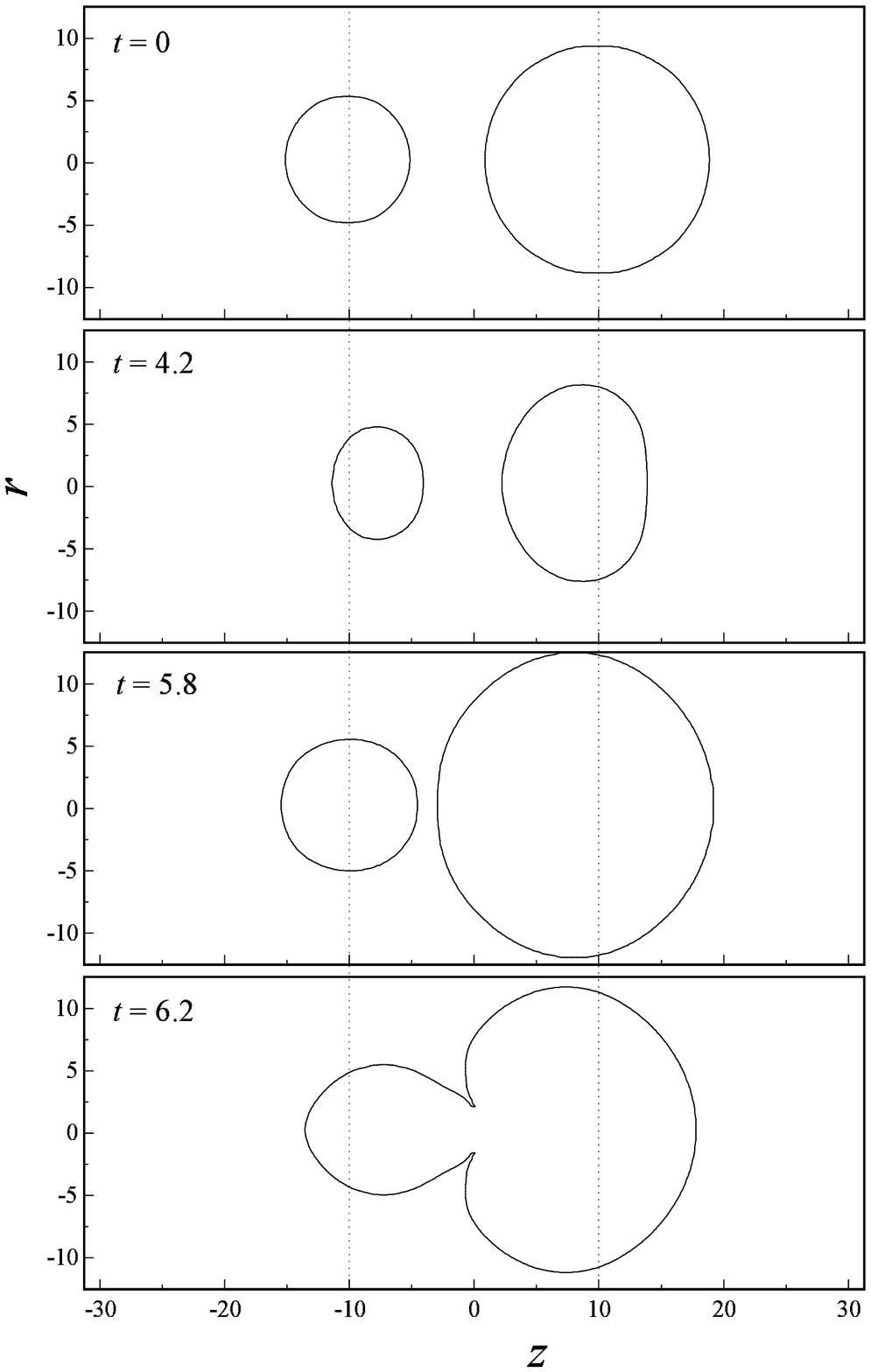}
\caption{
Calculated interfaces for $\omega  = \omega _{10} /1.8$ at selected 
times.}
\label{fig10}
\end{center}
\end{figure}

\begin{figure}
\begin{center}
\leavevmode
\epsfxsize = 8 cm
\epsffile{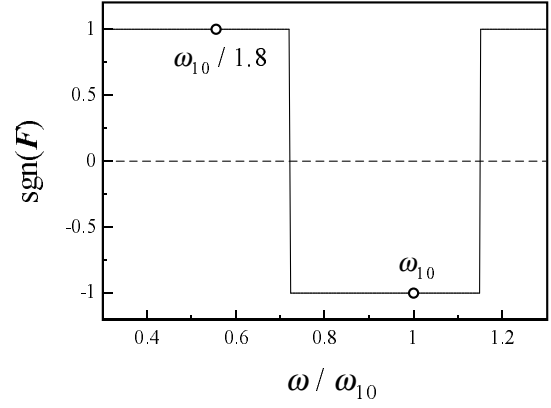}
\caption{
Theoretical result for the sign of the secondary Bjerknes force for 
$D = 20$ $\mu$m.}
\label{fig11}
\end{center}
\end{figure}

Figure \ref{fig8} shows the bubbles' mean radii and mass centers calculated for 
$\omega  = \omega _{10}$ and $\omega  = \omega _{10} /1.8$ 
($\approx \omega _{20}$), where $\omega _{j0}$ is the linear, monopole natural 
frequency of bubble $j$ determined by

\[
\omega _{j0}  = \sqrt {\left[ {3\gamma P_0  + (3\gamma  - 1)2\sigma /R_{j0} } \right]/\rho R_{j0}^2 } ,
\]
and the parameters for the MTS integration were set to $c1 = 0.2$, $c2 = 20$, 
and $c3 = 2$, resulting in $m3 \approx 8$. Also, Figs. \ref{fig9} and 
\ref{fig10} display the 
interfaces at selected times. We can observe the repulsion of the bubbles when 
$\omega  = \omega _{10}$, while the attraction when 
$\omega  = \omega _{10} /1.8$. These results are validated by the theory 
presented in Ref. \cite{ref51}, and reexamined in Refs. \cite{ref62,ref61}. 
The theory determines the sign of the interaction force (${\bf F}_{2B}$) by

\begin{equation}
\label{eq39}
{\mathop{\rm sgn}} ({\bf F}_{2B} ) = {\mathop{\rm sgn}} (\cos (\phi _1  - \phi _2 )) ,
\end{equation}
where

\[
\phi _1  = \tan ^{ - 1} \left( {B_1 /A_1 } \right) \in [0,2\pi ]
\]
with

\[
A_1  = \frac{{H_1 F + M_2 G}}{{F^2  + G^2 }} ,
\quad
B_1  = \frac{{H_1 G - M_2 F}}{{F^2  + G^2 }} ,
\]

\[
F = L_1 L_2  - \frac{{R_{10} R_{20} }}{{D^2 }}\omega ^4  - M_1 M_2 ,
\]

\[
G = L_1 M_2  + L_2 M_1 ,
\quad
H_1  = L_2  + \frac{{R_{20} }}{D}\omega ^2 ,
\]

\[
L_1  = (\omega _{10}^2  - \omega ^2 ) ,
\quad
L_2  = (\omega _{20}^2  - \omega ^2 ) ,
\]

\[
M_1  = \delta _1 \omega ,
\quad
M_2  = \delta _2 \omega ,
\]
where $\delta _j$, set to 
$(4\mu _l /\rho _l R_{j0}^2 ) + (\omega ^2 R_{j0} /c_l )$ in the present 
study, is the damping coefficient \cite{ref63}, $\mu _l$, $\rho _l$, and $c_l$ 
are the viscosity coefficient, the equilibrium density and the sound speed, 
respectively, of the liquid, and exchanging 1 and 2 (or 10 and 20) in the 
subscripts of these equations yields the expression for $\phi _2$. The positive 
sign of Eq. (\ref{eq39}) indicates attraction, while the negative indicates 
repulsion. Figure \ref{fig11} 
shows ${\mathop{\rm sgn}} ({\bf F}_{2B} )$ as a function of $\omega$ for 
$D = 20$ $\mu$m, revealing that the bubbles repel each other when 
$\omega = \omega _{10}$, while they attract when $\omega = \omega _{10} /1.8$. 
This theoretical prediction is well reproduced in the numerical 
results. (More detailed discussions from the physical viewpoint are given in 
Ref. \cite{ref64}.) Meanwhile, in the result for $\omega  = \omega _{10} /1.8$, 
we can observe the coalescence of the bubbles as a 
result of the attraction. This result proves the ability of the present 
methods to treat the topology change. 

\begin{figure}
\begin{center}
\leavevmode
\epsfxsize = 14 cm
\epsffile{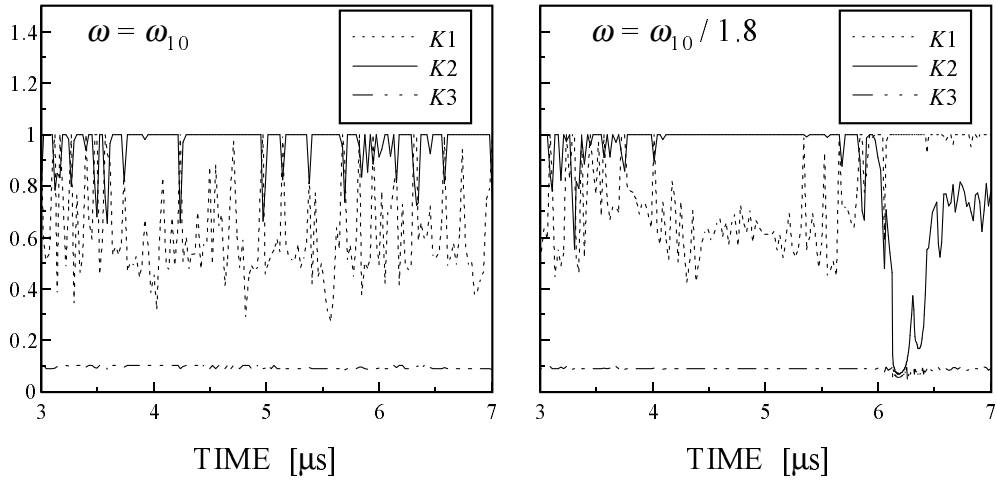}
\caption{
Time intervals normalized by $c1\,h/\max (\left| {\bf u} \right|)$ ($K$1), 
$c2\,h/\max (C_S )$ ($K$2), or $\Delta t_{st}$ ($K$3) as functions of time.}
\label{fig12}
\end{center}
\end{figure}

Figure \ref{fig12} shows $\Delta t$ normalized by 
$c1\,h/\max (\left| {\bf u} \right|)$, 
$c2\,h/\max (C_S )$, or $\Delta t_{st}$, as functions of 
time. This figure reveals that the fundamental time interval was mainly 
determined by the sound speed, whereas the flow velocity was responsible 
when a rapid motion of the interfaces, caused by the coalescence, occurred.

\begin{figure}
\begin{center}
\leavevmode
\epsfxsize = 12 cm
\epsffile{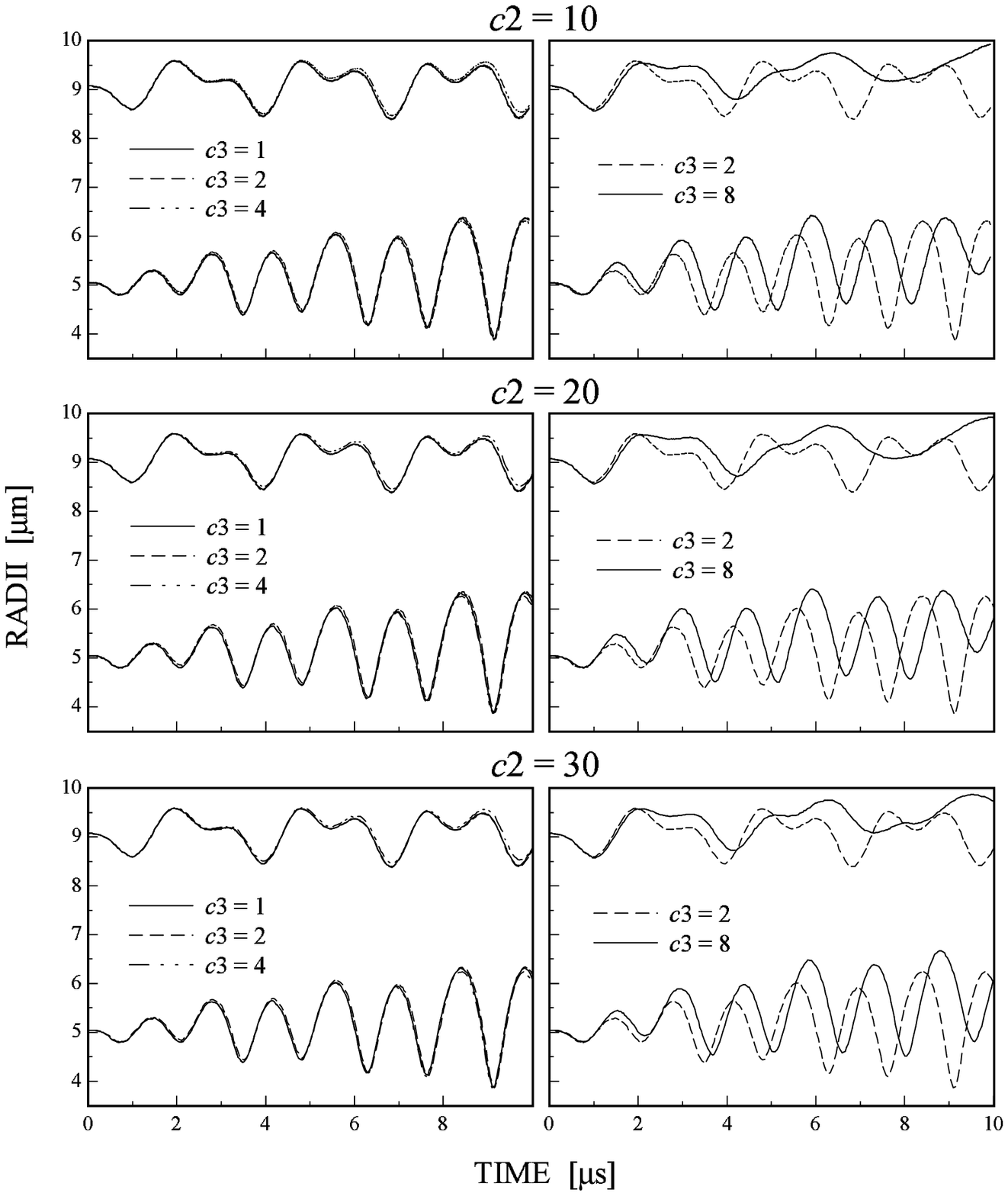}
\caption{
Bubbles' mean radii for $\omega  = \omega _{10}$ as functions of time, 
for $c1 = 0.2$ and different $c2$ and $c3$.}
\label{fig13}
\end{center}
\end{figure}

Using the example for $\omega  = \omega _{10}$, we investigate here the 
effectiveness of the MTS integration on this problem. Figure \ref{fig13} shows 
the bubbles' mean radii as functions of time, for $c1 = 0.2$ and different $c2$ 
and $c3$. Large differences cannot be seen between the results for $c3 \le 4$, 
while the results for $c3 = 8$ ($m3 \approx 2$) are obviously inaccurate. This 
result proves that the MTS integration has obviously contributed to the 
accurate solutions given for $c3 \le 4$.

During the computation for $\omega  = \omega _{10} /1.8$ introduced above, we 
sometimes observed negative $p$ near the 
interfaces. In such a case, we adopted $p = \max (p,\varepsilon _c P_0 )$, 
where $\varepsilon _c$ is set to $10^{ - 3}$; the numerical results were insensitive to the 
magnitude of this parameter. This problem would be overcome by employing the 
oscillation-free variants of the CIP \cite{ref65,ref66,ref67,ref68}.

\section{Conclusions}
\label{sec7}
In this series of articles, we have proposed an improved unified solver for 
compressible and incompressible fluids involving free surfaces, based on the 
CIP-CUP method, by adapting several improvements and modifications. (The 
most significant one given here in Part II is the adaptation of the MTS 
integration technique, which makes the determination of the time interval 
very flexible.) High accuracy and excellent robustness of the improved 
methods have been demonstrated by using examples of free-surface flows that 
contain both compressible and incompressible materials. The present methods, 
however, face the following challenges:

\bigskip

1. Optimizing the extrapolation function for the convection parts

In Part I of this series, we proposed five kinds of extrapolation functions 
used to solve the convection of the density. Although a concrete discussion 
has not been provided in the present paper, we observed that the most 
accurate result for a different problem was achieved by a different 
extrapolation function. The optimization of the function based on some 
criterion is, therefore, sorely expected.

\bigskip

2. Optimizing the time intervals

Although in the present study we have empirically determined the parameters 
for the MTS integration, an automatic, optimized determination should be 
useful in a practical application. The optimal time interval for the 
acoustic parts might be determined by a criterion based on the 
compressibility of fluids. If, for example, materials can be considered to 
be almost completely incompressible, as has been well known previously, one 
can use a time interval of an infinite CFL number with respect to the sound 
speed, whereas when a compressible material exists, as has been demonstrated 
in this paper, one needs a time interval depending on the sound speed. This 
means that the maximum values of $c2$ and $c3$ sufficient for accurate 
computations are rightly dependent on the compressibility.

\bigskip
\noindent
The above subjects will be addressed in a future paper.

\bigskip

\textbf{Acknowledgement}

The author would like to thank Dr. N. Masuda at Gunma University for his 
helpful supply of materials on the multi-time-step integration methods (or 
the individual timestep schemes) used in the astrophysical community.

\newpage
{\bf APPENDIX A: Averaging at phase boundary}

To discretize the acoustic, viscous, and surface tension terms on the 
staggered grids, we need the values of $\rho$ at the velocity positions, i.e., 
$\rho _{i + 1/2,j}$ and $\rho _{i,j + 1/2}$. In this appendix, we briefly 
discuss how to estimate them.

As discussed in Part I, the phase boundary is recognized using the zero 
level set of the ID function (Identification Function; the color or the 
level set function), $\phi$, and materials are identified using the sign of the 
function. If $\phi _{i + 1,j}  \cdot \phi _{i,j}  < 0$ is true, it is 
recognized that the call between ($x_{i + 1,j} ,y_{i + 1,j}$) and 
($x_{i,j} ,y_{i,j}$) contains an interface. In such a cell, we estimate the 
densities by the following VOF \cite{ref69}-like procedure, which implies the 
weighted linear interpolation:

\begin{equation}
\label{eq40}
\rho _{i + 1/2,j}^*  = \frac{{\left| {\phi _{i + 1,j}^{n + 1} } \right|\rho _{i + 1,j}^*  + \left| {\phi _{i,j}^{n + 1} } \right|\rho _{i,j}^* }}{{\left| {\phi _{i + 1,j}^{n + 1} } \right| + \left| {\phi _{i,j}^{n + 1} } \right|}}\quad {\rm for} \quad \phi _{i + 1,j}^{n + 1}  \cdot \phi _{i,j}^{n + 1}  < 0 ,
\end{equation}

\begin{equation}
\label{eq41}
\rho _{i,j + 1/2}^*  = \frac{{\left| {\phi _{i,j + 1}^{n + 1} } \right|\rho _{i,j + 1}^*  + \left| {\phi _{i,j}^{n + 1} } \right|\rho _{i,j}^* }}{{\left| {\phi _{i,j + 1}^{n + 1} } \right| + \left| {\phi _{i,j}^{n + 1} } \right|}}\quad {\rm for} \quad \phi _{i,j + 1}^{n + 1}  \cdot \phi _{i,j}^{n + 1}  < 0 ,
\end{equation}
where $\phi ^{n + 1}$ ($ = \phi ^*  = \phi ^{**}$. Note that the 
position of the interface is changed only when solving the convection 
parts.) is the ID function after solving the convection part. In the 
remaining cells, we use the simple average,

\[
\rho _{i + 1/2,j}^*  = \frac{{\rho _{i + 1,j}^*  + \rho _{i,j}^* }}{2} ,
\]

\[
\rho _{i,j + 1/2}^*  = \frac{{\rho _{i,j + 1}^*  + \rho _{i,j}^* }}{2} ,
\]
which has been used in the conventional CIP algorithm \cite{ref18}, or in 
others.

The above averaging scheme is also used to estimate the values of some other 
quantities at the velocity positions, such as the viscosity coefficient.

\newpage
{\bf APPENDIX B: On the surface tension term}

This appendix concerns how to calculate the surface tension term in our 
code.

In the model named the CSF (continuum surface force), proposed by Brackbill 
et al \cite{ref32}, the surface tension as a volume force is represented by

\begin{equation}
\label{eq42}
{\bf F}_{st}  = \sigma \kappa \nabla \theta ,
\end{equation}
where $\sigma$ is the surface tension coefficient, $\kappa$ defined as

\[
\kappa  = \nabla  \cdot \left( {\frac{{\nabla \theta }}{{\left| {\nabla \theta } \right|}}} \right)
\]
is the mean radius of the surface curvature, and $\theta$ denotes the color 
function defined as $\theta  = 1$ for a material, and $\theta  = 0$ otherwise. 
The singularity appearing in $\nabla \theta$ at the interface is mollified by 
some smoothing procedures \cite{ref32,ref33,ref70}. In the present study, the 
smoothed $\theta$ is constructed as follows:

\textbf{Step 1:} The non-smoothed color function is made from the ID 
function, which is an arbitrary function whose $\phi  = 0$ surface represents 
the interface \cite{ref19,ref42}, by setting $\theta _{i,j}  = 1$ for 
$\phi _{i,j}  > 0$ and $\theta _{i,j}  = 0$ for $\phi _{i,j}  < 0$. 
($\phi$ always has a non-zero value at the gird points \cite{ref19}.)

\bigskip

\textbf{Step 2:} The values of the color function at the velocity positions, 
$\theta _{i + 1/2,j}$ and $\theta _{i,j + 1/2}$, are determined by 
the averaging procedure described in Appendix A.

\bigskip

\textbf{Step 3:} Using the 4-point simple average, the weakly smoothed 
values of the color function at the original position ($i,j$) is obtained as

\[
\bar \theta _{i,j}  = 0.25(\theta _{i + 1/2,j}  + \theta _{i - 1/2,j}  + \theta _{i,j + 1/2}  + \theta _{i,j - 1/2} ) .
\]

\textbf{Step 4:} Using the following 5-point smoother, $\bar \theta$ is 
smoothed further:

\[
\bar \theta _{i,j}^{(m + 1)}  = (1 - \varepsilon )\bar \theta _{i,j}^{(m)}  + \varepsilon \frac{1}{4}(\bar \theta _{i + 1,j}^{(m)}  + \bar \theta _{i - 1,j}^{(m)}  + \bar \theta _{i,j + 1}^{(m)}  + \bar \theta _{i,j - 1}^{(m)} ) ,
\]
where $m$ ($ = 0,1, \cdots ,M - 1$) is the number of iteration, $M$ is the 
total number of iteration, and $\varepsilon$ is a small positive value.

Steps 2 and 3 are necessary to approximately take into account the phase 
property of the interface, which is not contained in $\theta$ given at Step 1, 
but is latent between the grid points. $\varepsilon$ and $M$ are typically 
set to $\varepsilon  = 0.15$ and $M = 20$.

Replacing $\theta$ with $\bar \theta ^{(M)}$, we calculate the 
surface-tension term by the second-order centered finite differencing on the 
staggered grids, as is done in Ref. \cite{ref32}.

\newpage

\begin{thebibliography}{}
\bibitem{ref1}
F. H. Harlow, A. A. Amsden, J. Comput. Phys. 8 (1971) 197.
\bibitem{ref2} 
R. I. Issa, J. Comput. Phys. 62 (1985) 40.
\bibitem{ref3}
T. Yabe, P. Y. Wang, J. Phys. Soc. Jpn 60 (1991) 2105.
\bibitem{ref4}
O. C. Zienkiewicz, J. Wu, Int. J. Numer. Meth. Eng. 35 (1992) 457.
\bibitem{ref5}
R. Klein, J. Comput. Phys. 121 (1995) 213.
\bibitem{ref6}
T. Schneider, N. Botta, K. J. Geratz, R. Klein, J. Comput. Phys. 155 
(1999) 248.
\bibitem{ref7}
R. Caiden, R. Fedkiw, C. Anderson, J. Comput. Phys. 166 (2001) 1.
\bibitem{ref8}
B. M\"{u}ller, in: Proc. 30$^{{\rm t}{\rm h}}$ Comput. Fluid Dynam. (Von 
Karman Institute for Fluid Dynamics, Lecture Series 1999-03), March 1999 
[review].
\bibitem{ref9}
T. Yabe, in: Proc. ECCOMAS 2000, Barcelona, September 2000 [review].
\bibitem{ref10}
T. Yabe, T. Mochizuki, H. Hara, in: Proc. LAMP '92, Nagoya. June 1992, 
p.387.
\bibitem{ref11}
T. Yabe, F. Xiao, D. Zhang, S. Sasaki, Y. Abe, N. Kobayashi, T. 
Terasawa, J. Geomag. Geoelec. 46 (1994) 657.
\bibitem{ref12}
T. Yabe, F. Xiao, J. Phys. Soc. Jpn. 62 (1993) 2537.
\bibitem{ref13}
Y. Zhang, T. Yabe, Comput. Fluid Dyman. J. 8 (1999) 378.
\bibitem{ref14}
T. Yabe, F. Xiao, T. Utsumi, J. Comput. Phys. 169 (2001) 556 [review].
\bibitem{ref15}
F. Xiao, J. Comput. Phys. 155 (1999) 348.
\bibitem{ref16}
S. Y. Yoon, T. Yabe, Comput. Phys. Commun. 119 (1999) 149.
\bibitem{ref17}
``\textit{The special issue of the CIP method''}, Comput. Fluid Dynam. J. 8 (1999).
\bibitem{ref18}
T. Yabe, T. Aoki, Comput. Phys. Commun. 66 (1991) 219.
\bibitem{ref19}
M. Ida, Comput. Phys. Commun. 132 (2000) 44.
\bibitem{ref20}
S. J. Aarseth, Mon. Notices Roy. Astron. Soc. 126 (1963) 223.
\bibitem{ref21}
J. Makino, Publ. Astron. Soc. Jpn. 43 (1991) 859.
\bibitem{ref22}
R. D. Swindoll, J. M. Haile, J. Comput. Phys. 53 (1984) 289.
\bibitem{ref23}
M. Tuckerman, B. J. Berne, G. J. Martyna, J. Chem. Phys. 97 (1992) 
1990.
\bibitem{ref24}
R. Zhou, E. Harder, H. Xu, B. J. Berne, J. Chem. Phys. 115 (2001) 2348.
\bibitem{ref25}
W. C. Chao, Mon. Wea. Rev. 110 (1982) 1603.
\bibitem{ref26}
T. Aoki, Comput. Fluid Dynam. J. 4 (1995) 279.
\bibitem{ref27}
A. A. Amsden, F. H. Harlow, LA-4370, Los-Alamos Scientific Laboratory 
(1970).
\bibitem{ref28}
S. Ito, in: Proc. The 43rd Nat. Cong. Theor. \& Appl. Mech., 1994, p. 
311 [in Japanese].
\bibitem{ref29}
S. V. Patankar, B. R. Baliga, Numer. Heat Transfer 1 (1978) 27.
\bibitem{ref30}
M. Ida, Y. Yamakoshi, Jpn. J. Appl. Phys. 40 (2001) 3846.
\bibitem{ref31}
M. Ida (unpublished).
\bibitem{ref32}
J. U. Brackbill, D. B. Kothe, C. Zemach, J. Comput. Phys. 100 (1992) 
335.
\bibitem{ref33}
M. Sussman, E. Fatemi, P. Smereka, S. Osher, Comput. Fluids, 27 (1998) 
663.
\bibitem{ref34}
P. Charrier, B. Tessieras, SIAM J. Numer. Anal. 23 (1986) 461.
\bibitem{ref35}
D. Mao, J. Comput. Phys. 92 (1991) 422.
\bibitem{ref36}
M. Ida, in: Proc. 10th Symp. Comput. Fluid. Dynam., Tokyo, Japan, 1996, 
p. 382 [in Japanese].
\bibitem{ref37}
M. Ida, in: Proc. 10th Comput. Mech. Conf., Tokyo, Japan, 1997, p. 17 
[in Japanese].
\bibitem{ref38}
D. H. Sharp, Physica D 12 (1984) 3 [review].
\bibitem{ref39}
J. Glimm, X. L. Li, R. Menikoff, D. H. Sharp, Q. Zhang, Phys. Fluids A 
2 (1990) 2046.
\bibitem{ref40}
D. Youngs, Phys. Fluids A 3 (1991) 1312.
\bibitem{ref41}
T. Yabe, H. Hoshino, T. Tsuchiya, Phys. Rev. A 44 (1991) 2756.
\bibitem{ref42}
W. Mulder, S. Osher, J. A. Sethian, J. Comput. Phys. 100 (1992) 209.
\bibitem{ref43}
S. K. Zhdanov, Physica D 87 (1995) 375.
\bibitem{ref44}
M. B. Schneider, G. Dimonte, B. Remington, Phys. Rev. Lett. 80 (1998) 
3507.
\bibitem{ref45}
L. Meignin, P. Ern, P. Gondret, M. Rabaud, Phys. Rev. E 64 (2001) 
026308.
\bibitem{ref46}
T. Yabe, T. Aoki, G. Sakaguchi, P. Y. Wang, T. Ishikawa, Comput. Fluids 
19 (1991) 421.
\bibitem{ref47}
M. Strasberg, J. Acoust. Soc. Am. 25 (1953) 536.
\bibitem{ref48}
A. Shima, Trans. ASME, J. Basic Eng. 93 (1971) 426.
\bibitem{ref49}
J. F. Scott, J. Fluid Mech. 113 (1981) 487.
\bibitem{ref50}
L. A. Crum, J. Acoust. Soc. Am. 57 (1975) 1363.
\bibitem{ref51}
E. A. Zabolotskaya, Sov. Phys. Acoust. 30 (1984) 365.
\bibitem{ref52}
J. R. Blake, P. B. Robinson, A. Shima, Y. Tomita, J. Fluid Mech. 255 
(1993) 707.
\bibitem{ref53}
P. A. Dayton, K. E. Morgan, A. L. Klibanov, G. Brandenburger, K. R. 
Nightingale, K. W. Ferrara, IEEE Trans. Ultrason. Ferroelect. \& Freq. 
Control 44 (1997) 1264.
\bibitem{ref54}
O. Louisnard, N. Lyczko, F. Espitalier, M. Urzedowski, Y. 
Vargas-Hernandez, C. Sanchez-Romero, Ultrason. Sonochem. 8 (2001) 183.
\bibitem{ref55}
C. Feuillade, J. Acoust. Soc. Am. 109, (2001) 2606.
\bibitem{ref56}
W. Lauterborn, T. Kurz, R. Mettin, C. D. Ohl, Adv. Chem. Phys. 110 
(1999) 295 [review].
\bibitem{ref57}
Z. Ye, A. Alvarez, Phys. Rev. Lett. 80 (1998) 3503.
\bibitem{ref58}
A. A. Doinikov, Phys. Rev. E 64 (2001) 026301.
\bibitem{ref59}
M. Ida, Phys. Lett. A 297 (2002) 210.
\bibitem{ref60}
M. Ida, J. Phys. Soc. Jpn. 71 (2002) 1214.
\bibitem{ref61}
M. Ida (submitted); e-print, physics/0109005.
\bibitem{ref62}
A. Harkin, T. J. Kaper, A. Nadim, J. Fluid Mech. 445 (2001) 377.
\bibitem{ref63}
A. Prosperetti, Ultrasonics 22 (1984) 69 [review].
\bibitem{ref64}
M. Ida, e-print, physics/0111138 (still in preparation).
\bibitem{ref65}
F. Xiao, T. Yabe, T. Ito, Comput. Phys. Commun. 93 (1996) 1.
\bibitem{ref66}
F. Xiao, Mon. Wea. Rev. 128 (2000) 1165.
\bibitem{ref67}
M. Ida, Comput. Fluid Dynam. J. 10 (2001) 159.
\bibitem{ref68}
M. Ida, Comput. Phys. Commun. 143 (2002) 142.
\bibitem{ref69}
C. W. Hirt, B. D. Nichols, J. Comput. Phys. 39 (1981) 201.
\bibitem{ref70}
M. Rudman, Int. J. Numer. Meth. Fluids 28 (1998) 357.

\end{thebibliography}
\end{document}